\documentclass[10pt,aps,pre,amssymb,twocolumn,groupedaddress]{revtex4-1}
\usepackage{graphicx}
\usepackage{bm}
\usepackage{amssymb}
\usepackage{amsmath}
\usepackage{color}
\newcommand{\q}{\theta}

\begin{document}
\title{Corner singularities and shape of stretched elastic sheets}
\author{Julien Chopin}
\affiliation{Instituto de F\'isica,  Universidade Federal da Bahia, Salvador-BA 40170-115, Brazil}
\author{Andreea Panaitescu and Arshad Kudrolli}
\affiliation{Department of Physics, Clark University, Worcester, MA 01610, USA}

\date{\today}
\begin{abstract}
We investigate the deformation of a longitudinally stretched rectangular sheet which is clamped at two opposite boundaries and free otherwise with experiments, numerical analysis and asymptotic analysis of the biharmonic elastic equation governing their planar equilibrium configurations. The displacement field of the sheet is measured  by tracking embedded fluorescent tracers with a digital image correlation (DIC) technique. The experiments and numerical finite element analysis (FEA) are found to be in overall good agreement except at the very corners where large deformations occur. We find that the deformed sheet can be broadly divided into a uniaxially stretched central region and two clamp dominated side regions.  A subregion characterized by a diverging stress can be identified at each of the four clamped-free corners within the clamp dominated region. We postulate that the divergence at the corners is regularized by nonlinear elastic deformations occurring in this subregion at the very corners and provide a nontrivial scaling for its size. Within the intermediate corner dominated subregion, measured displacements grow with distance $r$ from the corners as $r^{\alpha}$, with power $\alpha < 1$ consistent with the development of stress singularities at the intersection of the free and clamped edges. 
\end{abstract}

\maketitle

%%%%%%%%%%%%%%%%%%%%%%%%%%%%%%%%%%%%%%%%%%%%%
\section{Introduction} 
%%%%%%%%%%%%%%%%%%%%%%%%%%%%%%%%%%%%%%%%%%%%%
A stretched thin elastic sheet is a paradigm in understanding the shape and instabilities of highly deformable elastic solids~\cite{rivlin1951large,Cerda2003,Huang2010}. Rectangular sheets stretched longitudinally, while clamped at two opposite edges, contract transversely at the free edges due to Poisson's effect. As is by now well-known, wrinkles may occur for sufficiently thin sheets~\cite{Friedl2000,Cerda2003}, but their exact nature remains a matter of debate~\cite{Nayyar2011,Kim2012a,Healey2013,Taylor2014}. Such a boundary condition can further induce longitudinal wrinkles and creases when the sheet is twisted at low tension~\cite{Green1937,Chopin2013,Korte2011,Chopin2016}, and impact the wavelength of transverse wrinkles at high tension~\cite{Chopin2015,kudrolli2018tension}. But, even before the onset of instabilities, a stretched sheet embodies the complexity of thin sheet elasticity due to geometric nonlinearities even in Hookean (i.e. linear-elastic) materials. This may be one of the reasons why only crude approximations of the pre-wrinkling state have been considered to address their threshold and morphology~\cite{Cerda2003}. In the following, we focus on the deformation of a longitudinally stretched sheet to highlight the subtle interplay of elasticity and boundary conditions even in the absence of wrinkling.

Equilibrium configurations of an elastic sheet are governed by the biharmonic equation:
\begin{equation}
\Delta^2 \phi(x,y) = 0,
\label{eq:bihar}
\end{equation}
where $\Delta = \left(\partial^2/\partial x^2 +  \partial^2/\partial y^2\right)$, $x$ and $y$ are the coordinates along stretching and transverse direction, respectively, and $\phi(x,y)$ is the Airy stress function~\cite{landau1986theory}. Further,  $\phi(x,y)$ is defined through $\sigma_{xx} = \partial^2 \phi/\partial y^2$, $\sigma_{yy} = \partial^2 \phi/\partial x^2$, and $\sigma_{xy} = -\partial^2 \phi/\partial x\partial y$, where $\sigma_{ij}$ with $i,j = x,y$ are the 2D stress tensor components. The stress $\sigma_{ij}$ and strain $\varepsilon_{ij}$ are linearly related through the standard Hooke's law  under plane stress conditions~\cite{landau1986theory}. At the linear order in the displacement, the strain is defined through the displacement field $\textbf{u} = \left(u_x,u_y\right)$ as $\varepsilon_{xx} = \partial u_x/\partial x$, $\varepsilon_{yy} = \partial u_y/\partial y$, and $\varepsilon_{xy} = \left(\partial u_x/\partial y + \partial u_y/\partial x\right)/2$. Now, the boundary conditions imposed at an elastic sheet of length $L$ and width $W$ stretched by $\gamma_x = \Delta L/L$ by imposing a relative displacement $\Delta L$ between clamped edges reads, along the free edges ($0<x<L$): 
\begin{align}
\sigma_{yy}(x,0) = \sigma_{yy}(x,W) = 0,\label{eq:bc1}
\end{align}
and, along the clamped edges ($0<y<W$):
\begin{align}
u_{x}(0,y) = u_{y}(0,y) = 0,\label{eq:bc2}\\
u_{x}(L,y) = \Delta L \text{ and  } u_{y}(L,y)  = 0. \label{eq:bc3}
\end{align}

Harmonic and biharmonic equations with mixed boundary conditions, such as Eqs.~\ref{eq:bihar}-\ref{eq:bc3}, are encountered in various situations in physics including edge effect in electrostatics, wetting phenomena and crack propagation~\cite{eggers2015singularities}. In all these examples, the solution is singular as it exhibits a divergence at the location where the boundary condition changes abruptly. In the context of fracture mechanics, it is for example crucial to accurately predict the elastic field and the stress concentration developing near the tip of crack. Therefore, powerful methods have been developed to obtain the equilibrium configurations by solving Eq.~\ref{eq:bihar} with appropriate boundary conditions~\cite{muskhelishvili1953some,Leblond,broberg1999cracks,sih2013methods}. Yet, it remains a challenge to obtain analytical expressions over the entire domain. Progress can be made using asymptotic approaches allowing one to obtain the elastic field in a small region surrounding the crack tip. In this intermediate region, it is well-known in fracture mechanics that the elastic field exhibits a self-similar structure characterized by a universal exponent $\alpha = 1/2$~\cite{sun2012}. 

In a stretched elastic sheet, stress focusing and singularities also occur at the points where clamped edges meet free edges, as first discussed by Williams~\cite{Williams1952} and then by Bentham~\cite{Bentham1963}. In the case of a semi-infinite long sheet which is clamped at one end, Bentham showed~\cite{Bentham1963} that the longitudinal stress is singular and proportional to $ r^{-\alpha}$, where $r$ is the distance from the corner. But, unlike cracks, the exponent of the singularity $0 < \alpha < 1$ depends strongly on the Poisson ratio and geometrical factors. Later, Stern and Soni~\cite{stern1976computation} calculated the displacement fields and stress distribution which is also singular at the corner of a rectangular sheet of semi-infinite extent which is held under mixed free-clamped boundary conditions.  As far as we know, these calculated fields have not been tested quantitatively against experiments.
Further, because of the difficulty in deriving exact solutions of the equations for elastic sheets, finite element analysis (FEA) has been increasingly used in structural engineering to obtain stress and strain distributions. The efficacy of those techniques in cases where stress singularities can exist, as at the corners of clamped-free boundary conditions~\cite{Williams1952} remains unclear. 
Moreover, the overall shape and displacement field of a finite sized sheet has not been investigated in any detail. 

For materials with large Young's modulus such as steel and glass, the apparent change in shape of a finite sized sheet is not obvious for typical stresses applied before the material ruptures. In case of softer elastic materials such as latex rubber, the deformation becomes noticeable when examining images in the context of reports on wrinkling in  stretched thin sheets~\cite{Cerda2003,Nayyar2011,Healey2013}. Because a singularity is present at each of the four corners~\cite{Bentham1963}, their interactions can be further important to the observed displacement field of the finite-sized sheet and its edge shape. However, a study of the shape, which can further illuminate the stress distribution in the sheet as it is stretched, has not been reported.  This can not only lead to a better understanding of the large deformation regime in elastic materials but also the nature of instability observed in thin sheets under boundary forcing.
  
%%%%%%%%%%%%%%%%%%%%%%%%%%%%%%%%%%%%%%%%%%%%%%%%%%%%%%%%%%%%%%%%%%%%%%%%%%%%
\section{Protocols} 
%%%%%%%%%%%%%%%%%%%%%%%%%%%%%%%%%%%%%%%%%%%%%%%%%%%%%%%%%%%%%%%%%%%%%%%%%%%%
\begin{figure}
\begin{center}
\includegraphics{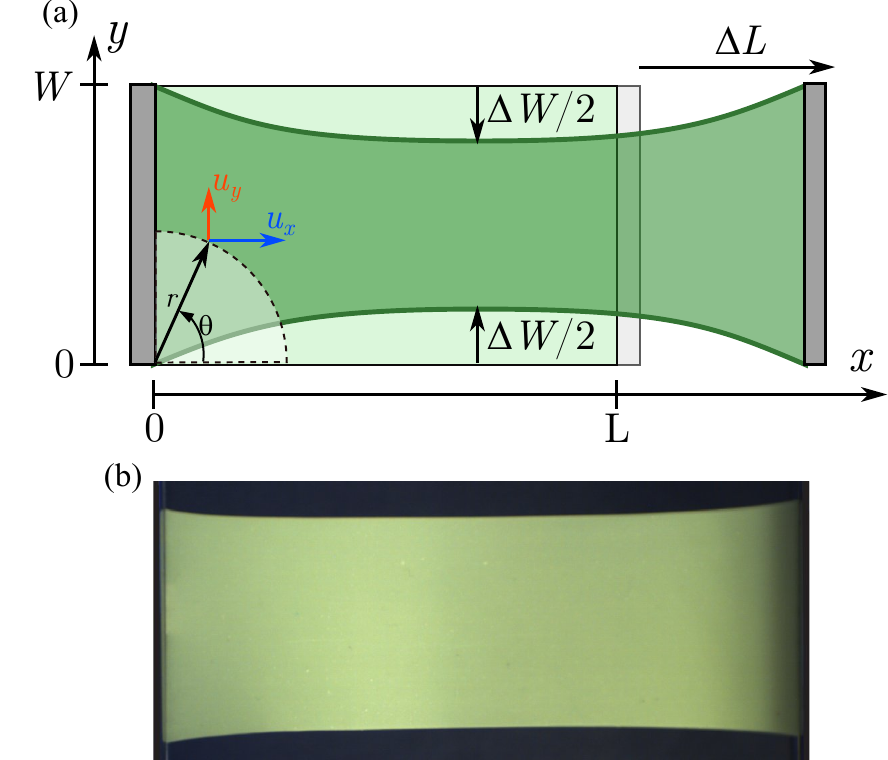}
\end{center}
\caption{(a) Schematic of the clamped stretched elastic sheet
and the coordinate system. (b) An image of a stretched
polyvinyl siloxane sheet ($L = 10$\,cm, $L/W = 2$, and $\gamma_x  = \Delta L/L = 0.35$).
}
\label{fig1}
\end{figure}

\subsection{Experimental System}
In order to investigate the deformation of elastic sheets, we experimentally study elastomers (polyvinyl siloxane and latex rubber) with initial length $L$, width $W$, and thickness $t$. The polyvinyl sheets were fabricated using molds in the laboratory, whereas the latex sheets were bought from a commercial supplier and ironed flat to remove any creases. The sheets were cut to a specified size before being mounted between two parallel flat aluminum clamps as shown schematically in Fig.~\ref{fig1}(a). The length and width of the sheet are aligned with the $x$ and $y$ axes of the Cartesian coordinate system with origin corresponding to the bottom left corner of the sheet as shown in Fig.~\ref{fig1}(a). The material properties of the sheets, including the Poisson ratio $\nu$ and the Young's modulus $E$, are obtained as part of the analysis of the displacement measurements discussed in the following sections.  

The elastic sheet is then stretched by moving one of the clamps which is attached to a motorized linear translating stage. Thus, the clamps stretch the sheet along the $x$ axis, while being held parallel to the $y$ axis. The resulting sheet stretched by a length increment $\Delta L$ is also shown  schematically in Fig.~\ref{fig1}(a), and an actual example corresponding to a polyvinyl sheet is shown in  Fig.~\ref{fig1}(b). Thus, the current sheet length is $L+\Delta L$ and the corresponding applied strain $\gamma_x = \Delta L/L$. Because of the  symmetry of the system, the resulting maximum transverse contraction of the width $\Delta W$ can be expected to occur at the mid-distance between the clamps. Thus, the corresponding width is $W-\Delta W$. % \sout{and the transverse strain $\gamma_y = -\Delta W/W$}. 
We also denote $u_x$ and $u_y$, the components of the displacement field along the $x$-axis and $y$-axis, respectively, in Fig.~\ref{fig1}(a). As shown in Fig.~\ref{fig1}(b), we image the sheets against a contrasting background with a digital camera with a resolution of $1824 \times 2048$ pixels to obtain the shape of the sheet using white light illumination. Thus, we can identify the edge of the sheet by processing the images using standard edge detection algorithms.

\begin{figure}
%\centering
\begin{center}
\includegraphics[width = 8.5cm]{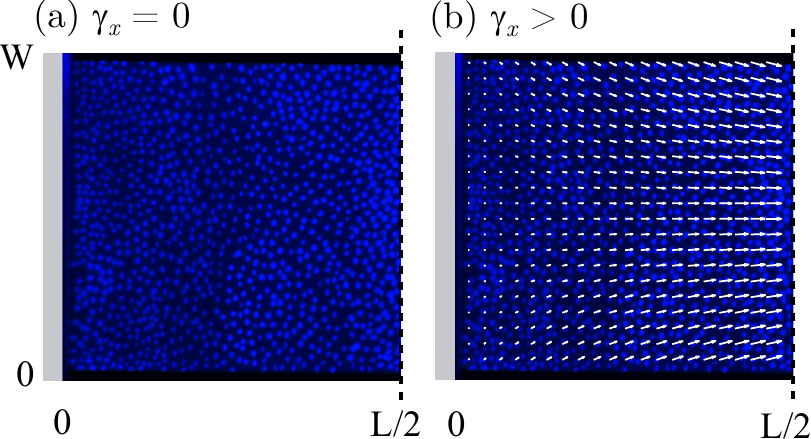}
\end{center}
\caption{(a) Image of a sheet with randomly placed markers which fluoresce under UV light with no applied strain $\gamma_x=0$. (b) Same image with the measured displacement field (white arrows) superimposed under an applied strain $\gamma_x = 0.0375$. Only the left half of the sheet is shown. ($L/W = 2$, $t/W = 1.9\times 10^{-3}$, $t=0.15$\,mm). 
}
\label{fig:dic}
\end{figure}

%%%%%%%%%%%%%%%%%%%%%%%%%%%%%%%%%%%%%%%%%%%%%%%%%%%%%%%%%%%%%%%
%\section{Displacement fields}
%%%%%%%%%%%%%%%%%%%%%%%%%%%%%%%%%%%%%%%%%%%%%%%%%%%%%%%%%%%%%%

%%%%%%%%%%%%%%%%%%%%%%%%%%%%%%%%%%%%%%%%%%%%%%%%%%%%%%%%%%%%%% Measurements by Digital Image Correlation
\subsection{Displacement Field with Digital Image Correlation}
\begin{figure}
\begin{center}
\includegraphics[width = 8.5cm]{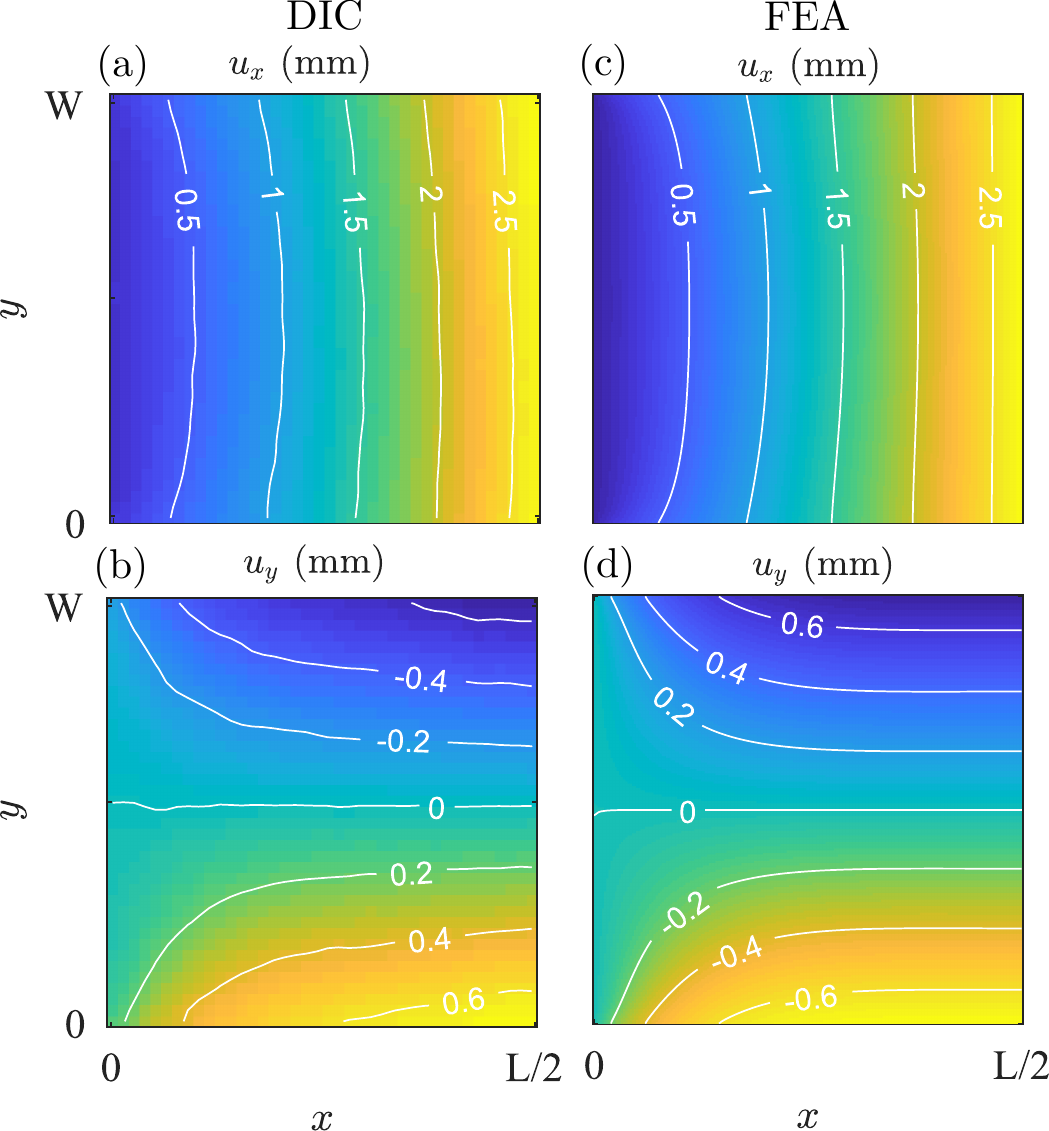}
\end{center}
\caption{Maps of the measured $u_x$ (a) and $u_y$ (b) in experiments corresponding to the displacement field shown in Fig.~\ref{fig:dic}(b). The iso-$u_x$ and iso-$u_y$ are also shown. Maps of $u_x$ (c) and $u_y$ (b) obtained with numerical FEA corresponding to the same system as in the experiments along with iso-$u_x$ and iso-$u_y$ contour lines. The experimental and numerical maps are observed to be overall in agreement.}
\label{fig:exptfield}
\end{figure}

The displacement field of the stretched sheet is measured using Digital Image Correlation (DIC) technique used to measure deformations in elastic medium~\cite{hild06}. This technique requires the surface to have a random texture. Because the surface of the materials used are featureless, we apply markers in a random pattern which fluoresce under UV light as shown in Fig.~\ref{fig:dic}(a). In this technique, the displacement field $\bm{u}(\bm{r})$ (with $\bm{r} = (x,y)$) is related to the intensity fields of the picture by $I_0(\bm{r})  = I_1(\bm{r} + \bm{u})$ where $I_0$ and $I_1$ are the intensity fields of the reference image before loading and the image after loading, respectively. The local displacement $\bm{u}$ of a material point originally at $\bm{r}$ is obtained by minimizing the cross-correlation function $C(\bm{u}) = \langle \left(I_1(\bm{r}+\bm{u})-I_0(\bm{r})\right)^2 \rangle$ with respect to $\bm{u}$ where $\langle ... \rangle$ is an average over a window centered on $\bm{r}$ with a lateral size which is typically between 8 to 64 pixels, depending on the image resolution. An interpolation of the pictures based on a Fourier decomposition along with a multiscale approach to locate the minimum of $C$ yield a robust and accurate calculation of the displacement with a subpixel resolution~\cite{hild06}. 

We first consider a latex sheet with size $L = 16$\,cm and $W = 8$\,cm, and thickness $t = 0.15$\,mm to investigate the observed displacement field as a function of applied uniaxial strain $\gamma_x$. Fig.~\ref{fig:dic}(b) shows an example of a displacement field superposed on the reference image when the elastic sheet is stretched by 6\,mm ($\gamma_x= 0.0375$). Because of the symmetry of the system, we focus on the left half of the system ($0< x < L/2$) to not only obtain the data with higher resolution but to show it with higher magnification as well. Here, it is to be noted that only a fourth of the vectors are shown for clarity, i.e. every alternate row and column is skipped to avoid overlaps in the plotting. Because of the size of the cross-correlation window used to detect displacements, the field is effectively averaged over an area of $1.3 \times 1.3$\,mm$^2$. This averaging allows for a high precision in the measurement of displacements of the order of 100\,$\mu$m,  corresponding to a resolution of approximately $1 \times 10^{-3}$ w.r.t. its length, which is of order of the thickness of the sheet. This small averaging window has little impact on the scale of the spatial trends of interest except at the very corners of the sheet.

\begin{figure*}
\begin{center}
\includegraphics[width = 17cm]{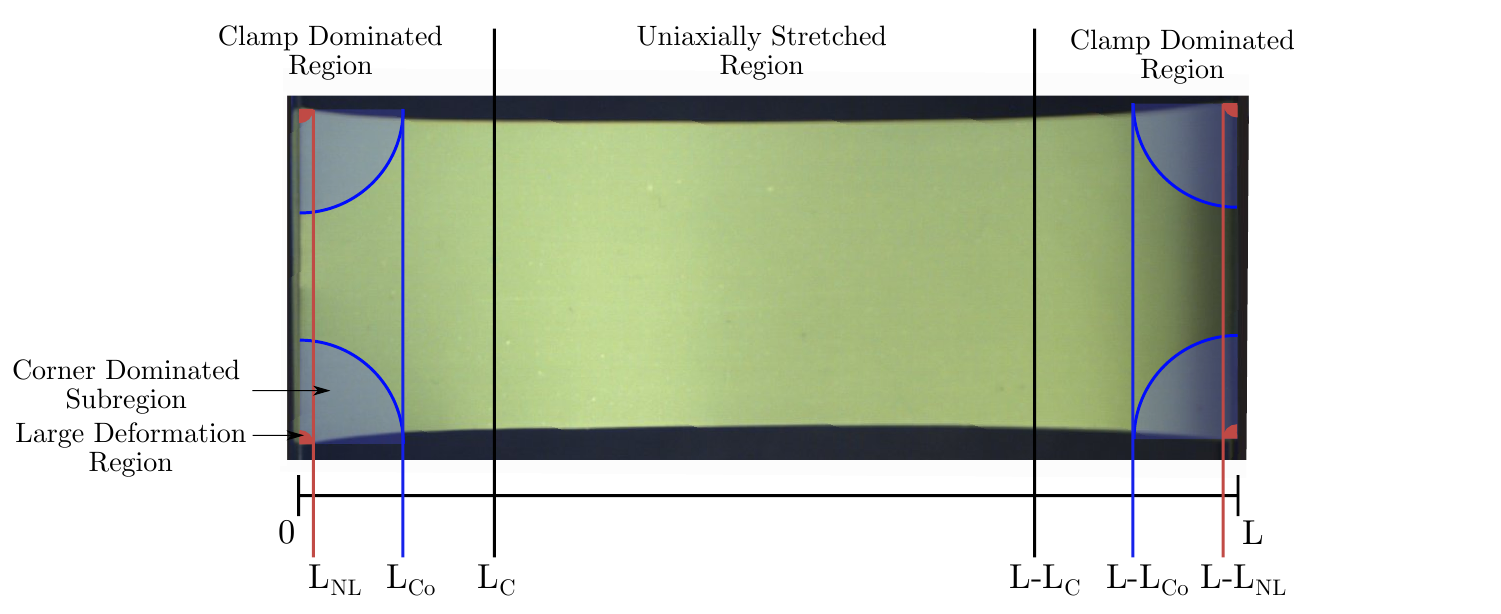}
\end{center}
\caption{A symmetric stretched sheet can be divided essentially into two main regions from the clamped boundaries. Further, two subregions can be identified as a function of distance $r$ from a corner. The corresponding distances along strain direction  $L_C$, $L_{Co}$, $L_{NL}$ are denoted. The corner dominated region corresponds to $L_{NL} <r <L_{Co}$.}
\label{fig:zones}
\end{figure*}

From Fig.~\ref{fig:dic}(b), one observes that the displacement field increases overall in magnitude from the left to the right because the left clamp is fixed and the right clamp (at $x = L$, not shown) is moved. Further, the direction of the arrows indicate that the relative magnitude of the displacement of the sheet in the longitudinal and transverse directions changes continuously with relatively smaller displacements in the transverse direction compared with the longitudinal direction.  

To see these trends in the displacement field more quantitatively, the map of $u_x$ is shown in Fig.~\ref{fig:exptfield}(a) along with iso-$u_x$ contour lines. The complementary map of $u_y$ is shown in Fig.~\ref{fig:exptfield}(b). We find that the central region of the sheet experiences a uniaxial stretching.  Near the clamp, one observes not only that the sheet displaces increasingly with $x$ but also somewhat more greatly at the sides compared with the center axis $y = W/2$ of the sheet. This is apparent from the curvature of the lines denoting the iso-$u_y$ contours in Fig.~\ref{fig:exptfield}(a), and the iso-$u_x$ contours in Fig.~\ref{fig:exptfield}(b). Thus, in this region near the clamps, the stretching is biaxial.

%%%%%%%%%%%%%%%%%%%%%%%%%%%%%%%%%%%%%%%%%%%%%%%%%%%%%%%%%%%%% Finite Element Analysis
\subsection{Finite Element Analysis}
We perform numerical analysis to further substantiate the experimental measurements. FEA has been used extensively to calculate the displacement fields and stress distributions in solids. Even though imposed mixed boundary conditions yields an elastic singularity at each corners of the stretched sheet, we find that Eqns.~\ref{eq:bihar}-\ref{eq:bc3} solved using the FEA module in MATLAB converges and yields stable results. 

The corresponding $u_x$ and $u_y$ maps are shown in Fig.~\ref{fig:exptfield}(c) and Fig.~\ref{fig:exptfield}(d), respectively, and are observed to be similar to those corresponding to the experiments.  In particular, the FEA displacement maps confirms the presence of a uniaxially stretched region and a clamp dominated region in the stretched elastic sheet with mixed clamped and free boundary conditions at opposite ends.

In the following sections, we will compare the obtained fields using the experimental and numerical methods quantitatively, besides analyzing the observations with asymptotic analysis. 

\begin{figure}
\begin{center}
\includegraphics{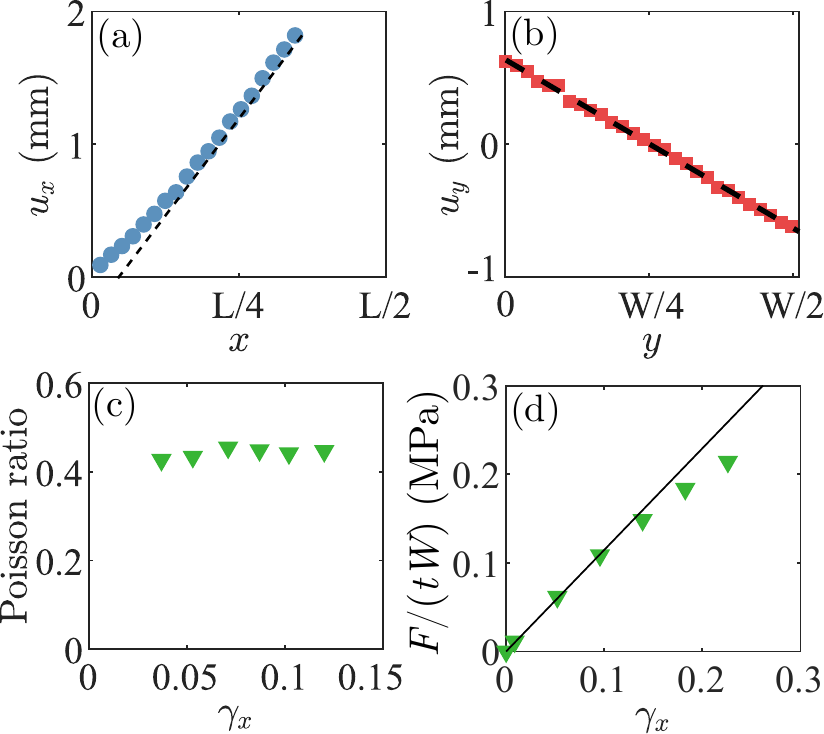}
\end{center}
\caption{(a) Displacement profile $u_x$ measured along the $x$ axis at $y = W/2$ for a latex sheet ($L/W = 2$, $t/W = 1.9 \times 10^{-3}$, $t = 0.15$\,mm). The clamp induces a slight deviation from linearity near $x=0$. A linear fit of the data away from the clamp gives $\varepsilon_{xx} = 0.0351$ and $u_0 = 0.25$\,mm (see Eq.\ref{eq:unistretch1}). (b) Displacement profile $u_y$ measured along the transverse direction at $x = 3L/4$ in the uniaxially stretched region. A linear fit of the data gives  $\varepsilon_{yy} =- 0.0159$. (c) The Poisson ratio $\nu = - \varepsilon_{yy}/\varepsilon_{xx}$ is essentially independent of the applied strain $\gamma_x$. (d) The nominal stress increases linearly with the strain for $\gamma_x <0.2$. A linear fit gives $E = 1.1$\,MPa.}
\label{fig:mechprop}
\end{figure}

%%%%%%%%%%%%%%%%%%%%%%%%%%%%%%%%%%%%%%%%%%%%%%% Stretching regimes
\section{Deformation regions in the stretched sheet}
From a visual inspection of the displacement field obtained by DIC and, further confirmed by FEA, we identify two regions: (a) a uniaxially stretched region, and (b) a clamp dominated region where the stretch is bi-axial as illustrated in Fig.~\ref{fig:zones}. The transition between these two regimes occurs at characteristic distance $L_C$ from the clamps. The clamp length $L_C$ may be expected to be dependent on the Poisson ratio, the sheet geometry and the applied strain.

As one attempts to understand the deformation of the sheet closer to where the free and clamped edges meet at the four corners in the region $x<L_C$, the displacement field may be characterized by universal features akin to singular field at the vicinity of a crack in a brittle material. Thus, we anticipate the existence of two other characteristics lengthscales, the corner length $L_{Co} (< L_C)$ and the nonlinear zone length $L_{NL} (< L_{Co})$ near each corner, as illustrated in Fig.~\ref{fig:zones}. For $r>L_{Co}$, the distance from the corner is too large for corner singularity to play a significant role. For $r<L_{NL}$, very close to the corner singularity, we expect that linear elasticity breaks down as is commonly observed in the process zone at the tip of a crack~\cite{broberg1999cracks}. As a consequence, one may postulate that universal features emerging from the stress singularity may develop over a distance $L_{NL} <r <L_{Co}$.

\begin{figure}
\begin{center}
\includegraphics[width = 8.5cm]{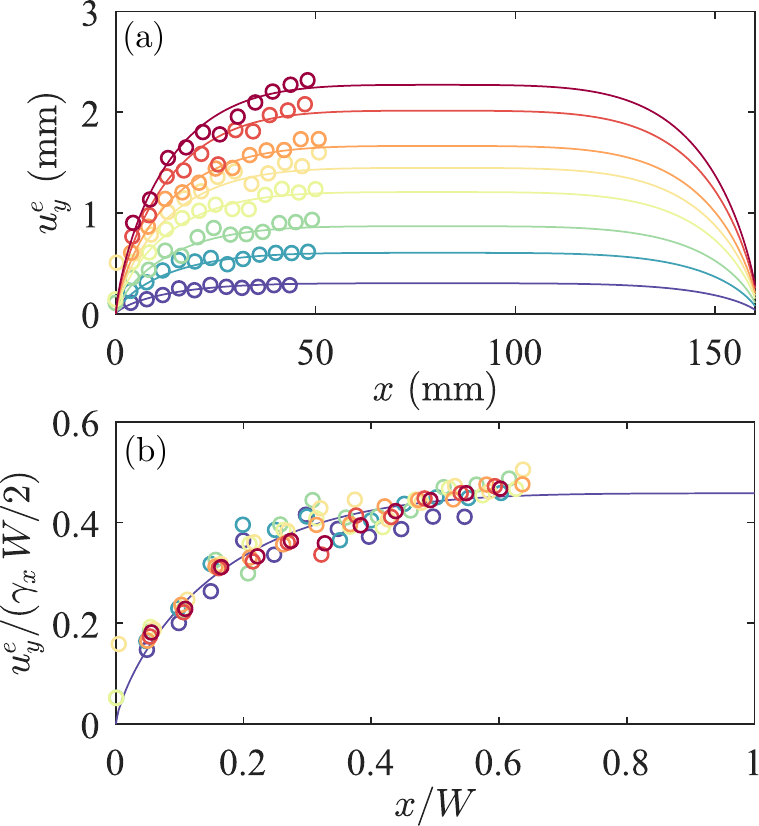}
\end{center}
\caption{(a) Measured transverse displacement $u_y$ normalized by $\gamma_x W/2$ for a latex sheet as a function of the normalized longitudinal coordinate $x/L$ for applied strain $\gamma_x$ in the range 0.09 to 0.30 ($L/W = 2$, $t/W = 1.9 \times 10^{-3}$, $t = 0.15$\,mm). (b) The data collapses onto a master curve independent of $\gamma_x$. The dashed line corresponds to the profile obtained by FEA with $\nu = 0.45$.}
\label{fig:uyprofile}
\end{figure}

%%%%%%%%%%%%%%%%%%%%%%%%%%%%%%%%%%%%%%%%%%%%%%%%%%%%%%%%%%%%%% Mechanical properties
%\subsection{Linear elastic properties}
\subsection{Uniaxially Stretched Region}
We now characterize the linear mechanical properties of the sheet from the DIC measurements in the uniaxially stretched region, \textit{i.e.} $L_C<x<L-L_C$. We integrate the transverse and longitudinal strains, $\varepsilon_{yy} = -\nu \varepsilon_{xx}$ and $\varepsilon_{xx}=\gamma_x$, respectively, in this region and obtain the displacement fields: 
\begin{align}
    u_x = u_0 + \gamma_x x,\label{eq:unistretch1}\\
    u_y = - \nu \gamma_x (y-W/2),\label{eq:unistretch2}
\end{align}
where, $u_0$ is a constant of integration which is dependent on the Poisson ratio and vanishes when $\nu = 0$.   From Eqns.~\ref{eq:unistretch1} and \ref{eq:unistretch2}, we can readily check that the iso-$u_y$ and iso-$u_x$ are parallel and perpendicular to the longitudinal edges, respectively, which is consistent with the contours shown in Fig.~\ref{fig:exptfield}.  In principle, $u_0$ is set by matching the solutions at the transition between the uniaxial and biaxial stretching regions. However, it is a priori unclear how to perform this in practice. 

In Fig.~\ref{fig:mechprop}(a), we plot the longitudinal displacement profile $u_x$ along the $x$ axis at $y=W/2$. A change of slope is observed near the clamp, signaling the transition to bi-axial strain. We then obtain the longitudinal strain measured as the slope of a linear fit of the data in the region where deformations are uniaxial. Using Eq.~\ref{eq:unistretch1}, we find $u_0 = 0.25$\,mm which is much smaller that the sheet length and, thus, it can be neglected to evaluate the local strain, $\varepsilon_{xx} \approx \gamma_x$. In Fig.~\ref{fig:mechprop}(b), we show the $u_y$ profile along the $y$ axis in the linearly stretched region. The displacements profile decreases linearly with $y$ with no observable effect at the boundaries. From the various applied strain $\gamma_x$, we measure the Poisson ratio from the relation $\nu = - \varepsilon_{yy}/\varepsilon_{xx}$, and find $\nu = 0.45 \pm 0.02$ (see Fig.~\ref{fig:mechprop}(c)). 

Finally, we measure the force $F$ for various applied strain. In Fig.~\ref{fig:mechprop}(d), a plot of the nominal stress $F/(Wt)$ as a function of the applied strain $\gamma_x$ reveals that the material is linear elastic for $\gamma_x <0.2$. For larger $\gamma_x$, strain softening develops as commonly observed in rubbers and elastomers~\cite{rivlin1951large}. In this regime, Young's modulus is $E = 1.1\,$MPa and the shear modulus defined as $\mu = E/(2(\nu+1)) = 0.38$\,MPa. No hysteresis or permanent deformation were observed upon unloading the sample for this entire range even in the non-linear regime.

%%%%%%%%%%%%%%%%%%%%%%%%%%%%%%%%%%%%%%%%%%%%%%%
\subsection{Edge profile and clamp length scale}
%%%%%%%%%%%%%%%%%%%%%%%%%%%%%%%%%%%%%%%%%%%%%%%
\begin{figure}
    \centering
        \includegraphics[width = 8.1cm]{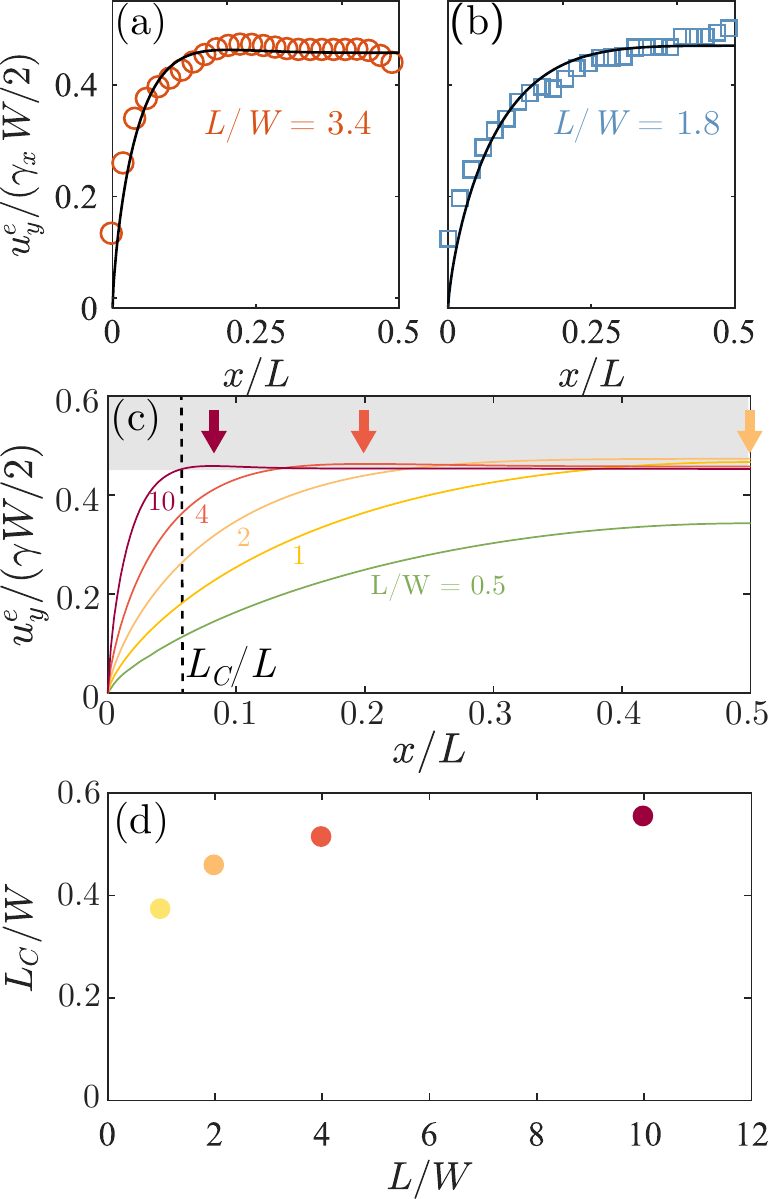}
    \caption{(a) The observed edge profiles for polyvinyl sheets with two different aspect ratios. ($L/W = 1.8$, $W = 6.3$\,cm, $t = 0.3$\,mm, and $\gamma_x = 0.25$) and ($L/W = 3.3$, $W = 3.1$\,cm, $t = 0.3$\,mm, and $\gamma_x = 0.30$). (b) Corresponding profiles obtained with FEA for various $L/W$ indicated are observed to show the same trends observed in the experiments. The arrows indicate the location of the shallow maximum in $u_y^e$. (c) The estimated $L_c$ increases with $L/W$ to approximately $0.6W$. }
    \label{fig:AR}
\end{figure}
We start by analyzing the edge profile $u^e_y$ of the sheet as a function of the applied strain and the aspect ratio $L/W$. This edge profile simply corresponds to the displacement field of the points at the free edge of the sheet, thus, $u^e_y(x) \equiv u_y(x,0)$, by definition. In Fig.~\ref{fig:uyprofile}(a), we show the edge profiles measured by DIC for $\gamma_x$ in the range $0.09$-$0.30$. The aspect ratio of the sheet is $L/W = 2$. At a fixed $\gamma_x$, the amplitude of the profile is increasing with $x$ until reaching a stationary value which indicates that the edges are parallel to the direction of stretch. This is consistent with the fact that the sheet is uniaxially stretched in that region. Thus, using Eq.~\ref{eq:unistretch2}, we have at $y=0$, $u_y^e = \nu \gamma_x W/2$ which explains the overall increase of the profile amplitudes with $\gamma_x$. Superposing experimental and numerical profiles obtained by FEA (solid lines) for the same $\gamma_x$, we observe good agreement, thus validating both the experimental measurements and numerical simulations.  

Further, as shown in Fig.~\ref{fig:uyprofile}(b), when normalizing the displacement by $\gamma_x W/2$, we observe a good collapse of the profiles on a master curve, not only in the central region, but also near the clamps over the entire range of applied strain. Thus, material nonlinearities can be neglected in the range of strain applied.

Next, we explore the role played by the aspect ratio at a fixed applied strain. Experimental data for two different aspect ratios are shown in Fig.~\ref{fig:AR}(a) and (b). We observe a good agreement with the numerical profiles calculated by FEA (solid lines). In Fig.~\ref{fig:AR}(c), we plot the normalized profile for aspect ratio in the range $0.5-10$ obtained using FEA. For aspect ratios $L/W < 1$, the profiles do not reach a stationary regime which indicates that, near the free edges, the stretching is bi-axial. As the aspect ratio is progressively increased, we observe the development of a flat central region when $W/L>1$.  Interestingly, the normalized profile reaches a maximum larger than $\nu$ which is the maximum value expected from Eq.~\ref{fig:uyprofile} as highlighted by the arrows in Fig.~\ref{fig:AR}(c). For $L/W \leq 2$, we find a unique, global maximum located at the center of the sheet ($x/L = 1/2$). But for $L/W > 2$, we observe the formation of two local maxima near the two opposite clamps (only the left half of the sheet is shown for clarity).  These qualitative features could not be observed in the experiments because of material and measurement uncertainties. 

To estimate the clamp length experimentally, $L_C$ is defined as the distance from the clamp where $u_y^e (L_C) = \nu \gamma_x W/2$ (a vertical dashed line is shown for the case of $L/W = 10$ in Fig.~\ref{fig:AR}(c)). This criterion corresponds to the point from where the free edges is essentially parallel. In Fig.~\ref{fig:AR}(d), we plot the measured $L_C$ as a function of the aspect ratio and find that for $L/W>1$, the clamp length normalized by the width is independent of $L/W$, and $L_C/W = 0.6 \pm 0.2$, where the variation occurs due to the method used to measure $L_C$.

\begin{figure}
    \centering
    \includegraphics[width = 7.5cm]{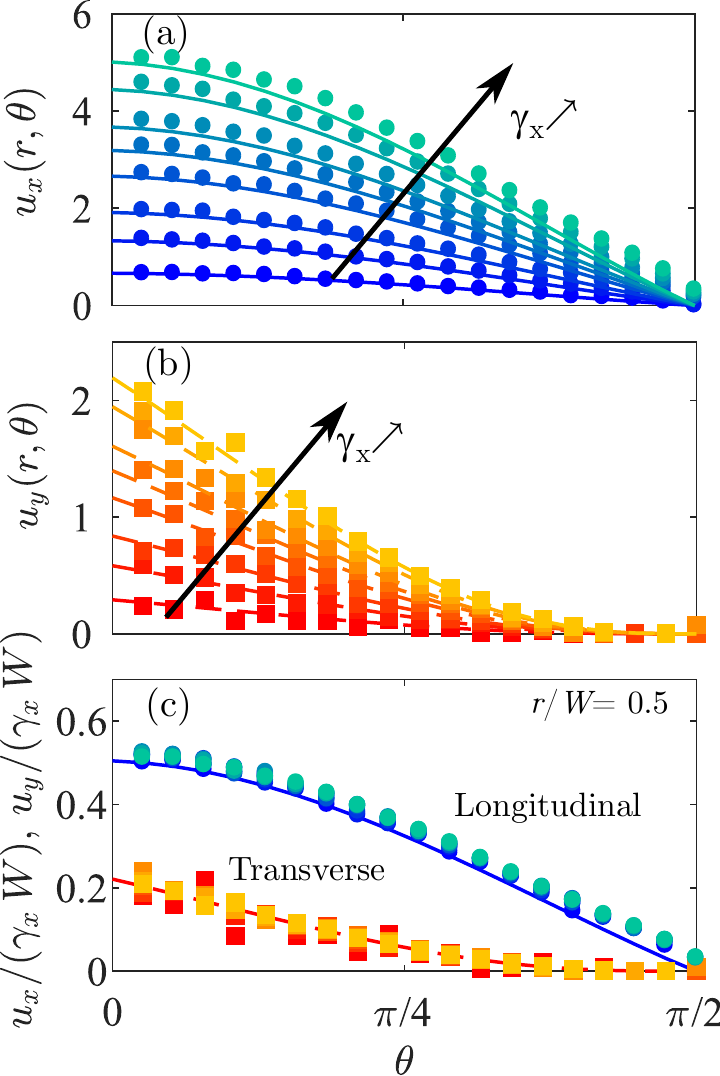}
    \caption{Longitudinal (a) and transverse (b) displacements profiles as a function of the angle varying $\gamma_x$ at a fixed distance from the corner $r/W = 0.5$. Solid lines are profiles obtained by FEA. (c) All the profiles collapse on two master curves in excellent agreement with numeric profiles obtained by FEA (solid lines).}
    \label{fig:AnalysiPol}
\end{figure}
\begin{figure}
    \centering
    \includegraphics[width = 6.5cm]{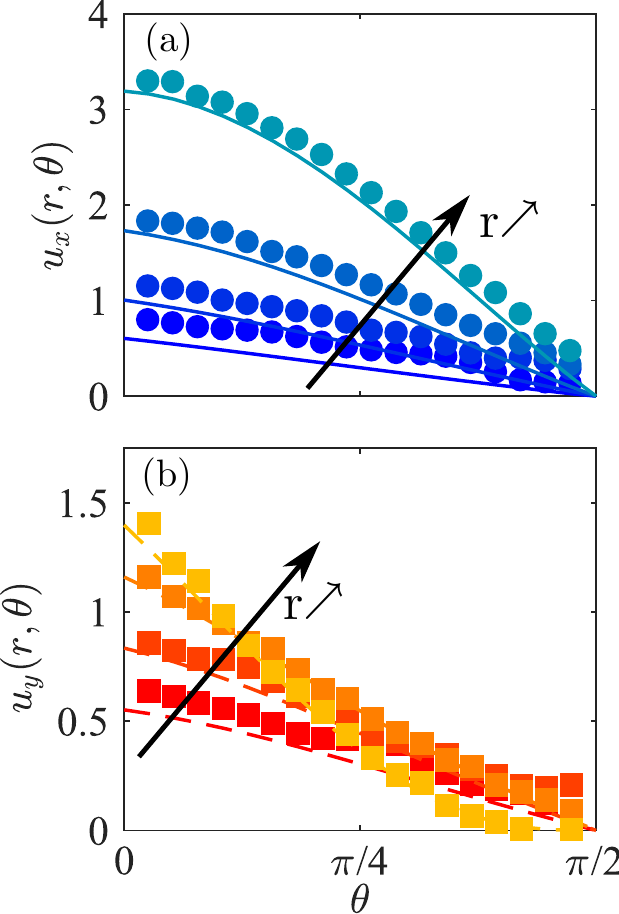}
    \caption{Longitudinal (a) and transverse (b) displacements profiles as a function of the angle varying the distance from the corner $r/W = 0.063,\,0.13,\, 0.25$, and 0.50, at fixed $\gamma_x = 0.079$. Solid lines are profiles obtained by FEA.}
    \label{fig:AnalysiPolRvaried}
\end{figure}

%%%%%%%%%%%%%%%%%%%%%%%%%%%%%%%%%%%%%%%%%%%%%% Displacement field in the clamped dominated region
\subsection{Displacement field in the clamp dominated region}
We now analyze the displacement field inside the sheet. It is convenient to use polar coordinates to express $u_x$ and $u_y$ (see Fig.~\ref{fig1}(a) for the notations). In Fig.~\ref{fig:AnalysiPol}(a) and (b), we plot the longitudinal and transverse displacement profiles obtained by DIC for $r/W = 0.5$ as a function of polar angle $\theta$. The different curves correspond to a fixed strain varied in the range $0.09 - 0.3$.  We observe that the overall profiles increase in amplitude with $\gamma_x$ and that they are smooth functions of $\theta$ monotonically decreasing to zero at $\theta = \pi/2$, thus satisfying the imposed boundary conditions at $x=0$. Excellent collapse of the profiles is obtained  when normalizing the displacement by $\gamma_x W$ as shown in Fig.~\ref{fig:AnalysiPol}(c). Comparing the measured profiles shown in Fig.~\ref{fig:AnalysiPol}(a-c) with numeric profiles obtained by FEA (solid lines), we obtain a quantitative agreement. 

In Fig.~\ref{fig:AnalysiPolRvaried}(a) and (b), we analyze the evolution of the displacement profiles increasing the distance from the corner keeping the applied strain fixed ($\gamma_x = 0.079$). We obtain a quantitative agreement between the profiles obtained by DIC  and  FEA  except very close to the corners corresponding to small $r/W$. At small distances, we find that the numerical results systematically underestimates the experimental results although the trend remains accurate. For the longitudinal displacement, the evolution of the overall amplitude is observed to increase monotonically with $r$. However, the transverse profiles shown in Fig.~\ref{fig:AnalysiPol}(b) reveal a more complex dependence with $r$ as the profiles overlap. This precludes a successful collapse of the data after a proper rescaling of the axis, at least in the relatively large range of $r$ considered. Hence, we consider a smaller subregion around the corners to develop further understanding of the observed displacement field near the clamps in the next section.

%%%%%%%%%%%%%%%%%%%%%%%%%%%%%%%%%%%%%%%%%%%%%%% Corner dominated region
\section{Corner dominated subregion}
We start by presenting the analytical results obtained by England~\cite{england1971stress}, and later by Stern and Soni~\cite{stern1976computation} using an asymptotic approach applied near the vicinity of the corner between semi-infinite clamped and free edges. 

\subsection{Asymptotic analysis}
The displacement components $u_x$ and $u_y$  near a corner upon loading can be decomposed as~\cite{stern1976computation}:
\begin{align}
u_x(r,\q) =  u_x^e(r)\,  \Phi_x(\theta) \left[1+ \mathcal{F}(r,\q)\right],\label{eq:ux}\\
u_y(r,\q) =  u_y^e(r)\,  \Phi_y(\theta) \left[1+ \mathcal{H}(r,\q)\right],
\label{eq:uy}
\end{align}
where, $u_x^e$ and $u_y^e$ are the longitudinal and transverse displacement at the edge, and $\Phi_x$ and $\Phi_y$ are smooth harmonic functions of $\q$ which depend on $\nu$. We explicitly include higher order contributions to the displacement field, $\mathcal{F}(r,\q)$ and $\mathcal{H}(r,\q)$.  These functions encapsulate the complex dependence of the displacement field with $r$, and are responsible for the absence of simple scaling law for $u_x$ and $u_y$ for arbitrary $r$ as evident in Fig.~\ref{fig:AnalysiPolRvaried}(b). However, in the asymptotic limit $r/W \ll 1$, $\mathcal{F}(r,\q)$ and $\mathcal{H}(r,\q)$ vanish yielding a separable form for $ u_x(r,\q)$ and $u_y(r,\q)$. In this limit, $u_x^e$ and $u_y^e$ reads~\cite{stern1976computation}:
\begin{align}
u_x^e(r) =  \frac{g_x(\nu)}{2 \mu } K_I r^{\alpha} \,\label{eq:uex}\\
u_y^e(r) =  \frac{g_y(\nu)}{2 \mu } K_I r^{\alpha}\,,\label{eq:uey}
\end{align}
where $K_I$ is the stress intensity factor in mode I (traction mode), $g_x(\nu)$ and $g_y(\nu)$ are smooth functions of $\nu$, and $\mu = E/[2(1+\nu)]$  is the shear modulus. Further, $\Phi_x(\theta)$ and $\Phi_y(\theta)$ are monotonically decreasing functions with $\Phi_x(0) = \Phi_{y}(0) = 1$, and $\Phi_x(\pi/2) = \Phi_{y}(\pi/2) = 0$. The expression for $g_x(\nu)$, $g_y(\nu)$, $\Phi_x(\theta)$, and $\Phi_y(\theta)$ can be found in Appendix~\ref{apend}.
The exponent $\alpha$ is the solution of~\cite{stern1976computation}:
\begin{equation}
    \cos \alpha = \frac{2\alpha^2}{\kappa}-\frac{\kappa^2+1}{2 \kappa}\,,
\label{eq:lambda}
\end{equation}
where $\kappa = (3-\nu)/(1+\nu)$ in case of plane stress. For an incompressible material $\nu = 0.5$, and thus $\kappa = 5/3$, $\alpha = 0.69 \approx 2/3$, and $\mu = E/3$. In the case of the materials used in our experiments where $\nu = 0.45$, we have $\alpha = 0.71$.  Because the amplitude of the displacement grows sublinearly with $r$, the stress is singular at the four corners. Indeed, we have $\sigma_{rr} \sim \mu \partial u_r/ \partial r$, and thus $\sigma_{rr}  \sim K_I r^{\alpha -1}$, with $\alpha< 1$, which diverges as $r$ goes to zero. This is obviously unphysical as was already noted by Williams~\cite{Williams1952}. Therefore, it is necessary to introduce a cutoff length at a small scale below which the asymptotic solution is no longer valid.

The functional dependence of the stress intensity factor with the applied loading, mechanical properties, and the sheet dimensions has not been yet given. By dimensional analysis, we expect:
\begin{equation}
    K_I = k_I(\nu) \frac{F}{A} W^{1-\alpha}\,,
\label{eq:KI}
\end{equation}
where $k_I$ is a smooth function of $\nu$. As the Poisson ratio $\nu$ tends to zero, the stress field is uniaxial over the entire length of the sheet, without any bi-axial stress building up in the clamp dominated region. Therefore, we expect that clamp dominated region to disappear altogether with the elastic singularity. Then, we argue that, as $\nu \rightarrow 0$, the extension of the clamp dominated region $L_C$ scales as $L_C \sim W \nu^{\gamma}\, (\gamma>0),$ and the strength of the singularity scale  $K_I \sim k_I \sim \nu^{\beta}\, (\beta>0).$ 

\begin{figure}
    \centering
    \includegraphics[width = 8.5cm]{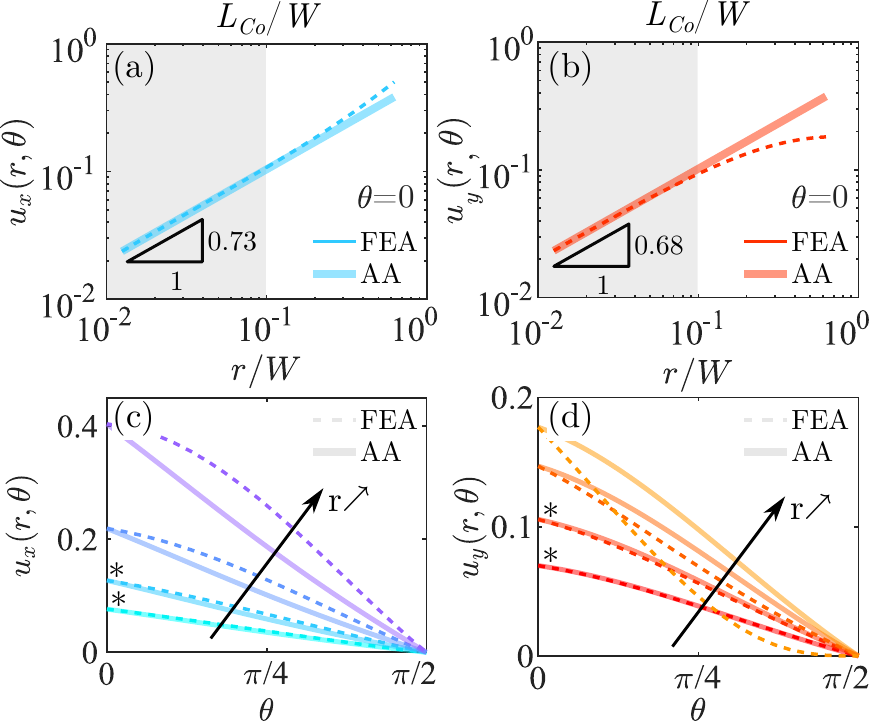}
    \caption{The dependence of the longitudinal displacement $u_x$ (a) and transverse displacement $u_y$ (b) with distance from the corner $r$ for $\theta = 0$. Good agreement between numerics (dashed lines) and asymptotic analysis (solid lines) is obtained for $r/W < L_{Co}/W = 0.1$. Profiles are obtained by FEA (dashed lines) and asymptotic analysis (solid lines).   $\theta$-dependence of $u_x$ (c)  and $u_y$ (d) for $r/W = $ 0.063, 0.1, 0.25, and 0.5. Good agreement between numerics (FEA, dashed lines) and asymptotic analysis (AA, solid lines) is obtained for $r < L_{Co} = 0.1\,W$.}
    \label{fig:FEA_Stern}
\end{figure}

%%%%%%%%%%%%%%%%%%%%%%%%%%%%%%%%%%%%%%%%%%%%%%%%%%%%%%%%%%%%%%%%%%%%%%%
\subsection{Range of validity of the asymptotic displacement field}
As argued earlier, the asymptotic solutions derived within a linear elasticity framework are expected to be accurate over a finite range of distance from the corner, bounded from below by the large deformation or nonlinear zone, and from above by the fact that the sheet is finite sized. Therefore, to compare the asymptotic solution with experiments, it is crucial to address its range of validity and evaluate the cutoff lengths at small and larger scales which can depend on the Poisson ratio, sheet geometry and applied strain. 

Now, the higher order terms in the Eqns.~\ref{eq:ux} and~\ref{eq:uy} become of the same order as the singular terms, far enough from the corner. Since analytic expressions for the higher order terms are not available, we resort to discussing the range of validity of the asymptotic predictions based on the numerical results. The longitudinal and transverse displacements along the direction $\theta = 0$ are shown in a log-log plot in Fig.~\ref{fig:FEA_Stern}(a) and Fig.~\ref{fig:FEA_Stern}(b), respectively. We define the corner length $L_{Co}$ as the distance below which the numeric profile and asymptotic prediction are in agreement to within 10\% relative difference. With this definition, we obtain a measure of the corner length, $L_{Co} = 0.1\,W$. For $r<L_{Co}$, a fit of the data gives $u_x(r,0) \sim r^{0.73}$ and $u_y(r,0) \sim r^{0.68}$, in good agreement with prediction of $\alpha = 0.71$ in case where $\nu = 0.45$ by Stern and Soni~\cite{stern1976computation}. 

Next, we analyze the angular dependence of the displacement field. In Fig.~\ref{fig:FEA_Stern}(c) and (d), we plot the longitudinal and transverse displacement as a function of $\theta$  for $r/W$ in the range $0.0625-0.5$. Consistent with previous comparisons, we obtain good agreement between numerics (dashed lines) and asymptotic analysis (solid lines) for $r/W < L_{Co}/W = 0.1$, corresponding to the profiles with a star label.

\subsection{Small scale cutoff}
Now, we focus on the small scale cutoff below which the model equations used cannot be expected to be valid. It has been long realized~\cite{Williams1952} that the divergence of the stress at the singularity is not physical and is regularized by nonlinear processes which operate in any material at sufficiently large stresses. Such processes mean that the linear elastic equations used are no long valid. For example, in fracture mechanics, which is governed by similar elastic equations~\cite{broberg1999cracks,Leblond,sun2012}, a {\it fracture process zone} is well known to occur near the tip of a fracture due the singular nature of the solutions there. The nature of the nonlinear processes and the size of the process zone are highly dependent on the type of material, the type of loading and the geometry of the sample. In our experiments, we did not observe irreversible deformation due to plasticity or damage with the material used over the range of applied strain. Thus, nonlinear elastic (reversible) deformations alone may suffice to regularize the singularity at the corner. 

Assuming that elastic nonlinearity becomes dominant for stresses $\sigma \sim \mu$ or strain of $\mathcal{O}(1)$, we can provide a natural lengthscale $L_{NL}$ below which linear elasticity breaks down. From Eqs.~\ref{eq:uex} and \ref{eq:uey}, the typical strain experience at a distance $r$ is given by $\varepsilon \sim K_I/\mu r^{\alpha - 1}$. Thus, when $r \approx L_{NL}$, $K_I/\mu L_{NL}^{\alpha -1}\sim 1$, yielding $L_{NL} \sim \left(K_I / \mu\right)^{1/(\alpha-1)}$. Using Eq.~\ref{eq:KI}, we obtain: 
\begin{equation}
    L_{NL}/W = f(\nu)\, \gamma_x^{1/(1-\alpha)},
\end{equation}
where, $f(\nu)$ is an unknown function of $\nu$ which should vanish as $\nu \rightarrow 0$. Similar scaling for the extent of the process zone, where nonlinearities dominate, has been derived in the context of crack propagation in gels and adhesives~\cite{baumberger2008magic,bouchbinder2014dynamics,creton2016fracture}. With the typical strain and width used in the experiment ($\gamma_x \sim 0.1$, $W = 80$\,mm), and assuming that $f$ is $\mathcal{O}(1)$ for incompressible material, $L_{NL}/W \sim 10^{-3}$, thus $L_{NL}$ is of order 0.1\,mm. This length scale is of order of the spatial accuracy of the DIC technique and the thickness of the sheet, and cannot be clearly resolved in our experiments. 

%%%%%%%%%%%%%%%%%%%%%%%%%%%%%%%%%%%%%%%%%%%%%%%%%%%% Experimental results
\subsection{Comparisons with DIC and FEA methods}
\begin{figure}
\begin{center}
\includegraphics[width = 8.5cm]{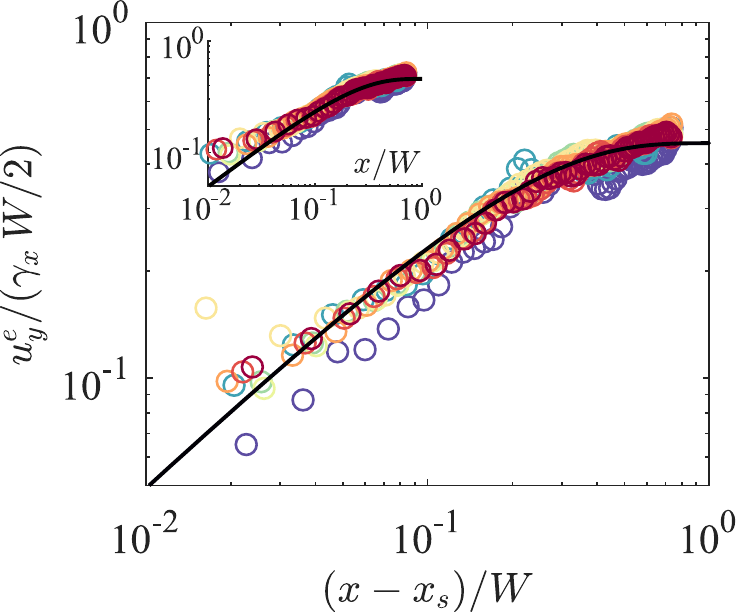}
\end{center}
\caption{Normalized edge profile from experiment (colored disks) and numeric simulations (solid line) as a function of $x/W$. Measured profile exhibits a power law compatible with numeric profiles and theoretical predictions introducing an offset $x_s /W = -0.01$. Insets : same data without offset.}
\label{fig:edgeloglog}
\end{figure}
\begin{figure}
\begin{center}
\includegraphics{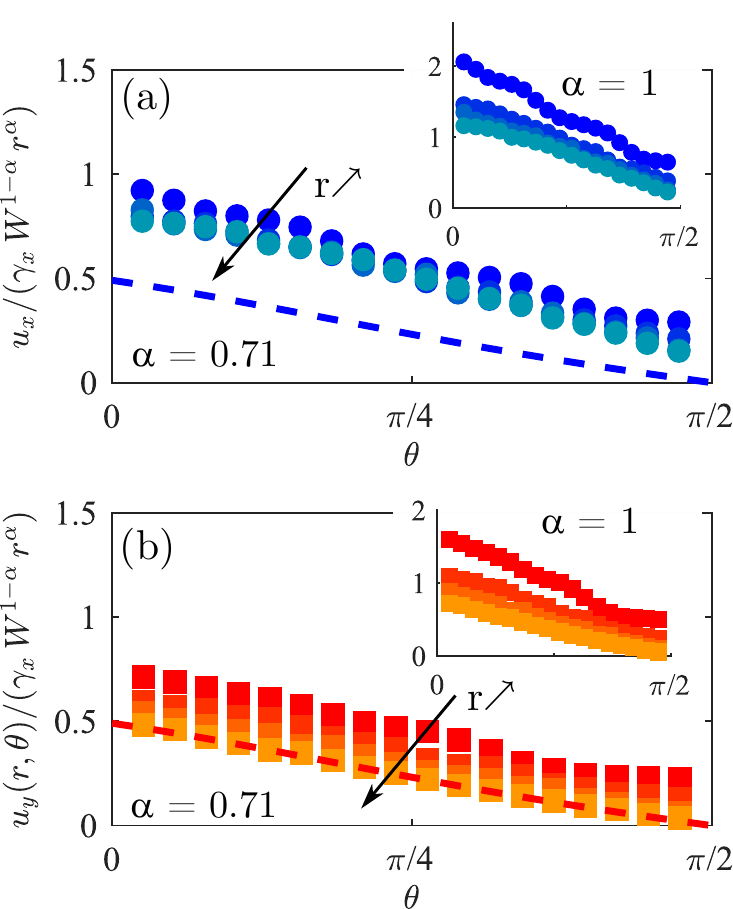}
\end{center}
\caption{Displacement profiles in the longitudinal (a) and transverse (b) direction measured at a distance $r = 5,\, 10,\,15,\,20$\,mm from the corner ($\gamma_x = 0.09$). A good collapse of the data is obtained when the displacements are normalized using $\alpha = 0.71$ rather than $\alpha = 1$ (see inset).}
\label{fig:exptsim}
\end{figure}
We now compare the results obtained from experiments and simulations with the predictions from the asymptotic analysis having obtained the inferred range of $L_{NL}<r<L_{Co}$, where $L_{NL}/W = 10^{-3} \sim 0$ and $L_{Co}/W = 0.10$. In Fig.~\ref{fig:edgeloglog}, we show in a log-log plot, the transverse displacement at the edge $u_y^e$ normalized by $\gamma_x\,W/2$ as a function of $x/W$ (see inset). We find that the experimental profiles are compatible with the theoretical prediction and FEA profiles after introducing a small offset $x_s/W = -0.01$ in the $x$-axis. Because edge profile develops a power behavior over a range of lengthscales which are difficult to reach experimentally, the agreement with theory is very sensitive to an offset in the origin of the $x$ axis.  

Next,  we compare the $\theta$-dependence of experimental displacements field with theoretical predictions for $r/W =$ 0.0625, 0.10, 0.2, and 0.25 in Fig.~\ref{fig:exptsim}. For $r/W < L_{Co}/W = 0.1$, we expect that the experimental profiles measured at various distance from the corner collapse on a single power law when normalized by $ r^{\alpha}$ with $\alpha = 0.71$. In Fig.~\ref{fig:exptsim}(a),  we plot the longitudinal displacement normalized by $\gamma_xW^{1-\alpha}r^{\alpha}$ ($\sim r^{\alpha}$) as a function of $\theta$. We observe a good collapse of the data when $\alpha = 0.71$. For comparison, in the inset, we show the effect of a rescaling using $\alpha = 1$. Although the evolution of the profile is consistent with numeric profiles obtained by FEA (dashed line), the experimental profiles are shifted upward by an overall constant.  Then, we show the result of the same analysis for the transverse displacement in Fig.~\ref{fig:exptsim}(b). We observe a better collapse of the data using $\alpha = 0.71$ than $\alpha = 1$ (see inset), confirming that the distribution of the displacement field. 

Thus, we clearly find that the strain and stress fields are controlled by a nontrivial exponent. Although the experimental data do not allow a precise measure of the exponent of the singularity, we demonstrate unambiguously that it is smaller than unity confirming the existence of an elastic singularity at the corners of a longitudinally stretched sheet.

%%%%%%%%%%%%%%%%%%%%%%%%%%%%%%%%%%%%%%%%%%%%%%%%%%%% 
\section{Conclusions} 
%%%%%%%%%%%%%%%%%%%%%%%%%%%%%%%%%%%%%%%%%%%%%%%%%%%%

In summary, we obtain the displacement field and the shape of elastic sheets under longitudinal stretching with complementary experimental and theoretical models. In particular, we accurately measure the local displacements experienced by the sheet using DIC techniques over a wide range of applied strains. We also calculate the displacement field using finite element analysis with equations used to model planar deformation of sheets under boundary applied stresses. This numerical analysis is performed with the same sized sheets as in the experiments, but assuming that the thickness of the sheet does not evolve, even when a significant range of strain is applied. Nonetheless, we find that linear elasticity provides a reasonable description of the overall displacement field and shape of the free edges for applied strained $\gamma_x$ at least up to 0.3. 

While a reasonably simple form of the elastic equations are found to describe the overall deformations, these equations cannot be fully solved analytically. Thus, to make progress toward developing an analytical framework to understand the observed sheet displacement field, we identify regions and subregions in the deformed sheet which are amenable to further analysis. We find that the sheet can be divided overall into a simple uniaxially stretched central region, and a more complex clamp dominated region near each clamped edge. The length scale from the clamps where linear uniaxial stretching occurs can be understood as resulting from a trade off between satisfying the no slip boundary condition imposed at the clamps and minimizing the additional elastic energy induced by the transverse stresses. 

Further, inside the clamp dominated regions, we identify two subregions depending on the distance from each clamped-free corner of the sheet. In the corner dominated subregion, we find that asymptotic analysis, performed in the context of sheets with semi-infinite clamped and free edges under strain by Stern and Soni~\cite{stern1976computation}, is found to be in excellent agreement with FEA analysis both for the evolution of the displacement field with distance from the corner ($\sim r^{\alpha}$ with $\alpha = 0.71$ for Poisson ratio $\nu = 0.45$) and the angular dependence. The extent of this subregion characterized by a diverging stress is found to be of order $L_{Co} = 0.1\, W$ for the elastomers used. We show that asymptotic and FEA analysis give a reasonable description of the measured displacement fields obtained by DIC technique.  While the power law $u_y \sim r^{\alpha}$ could not be checked experimentally in terms of distance $r$ because of a lack of sufficient range of length scales within the corner dominated zone, we observe good agreement between the asymptotic solutions and the experiments for the angular dependence. Further, we infer that the derived power law of $0.71$ is consistent with our measurements. Very close to the corners, we postulate the existence of a large deformation subregion where nonlinear process may occur because of the large diverging stresses there, analogous to the process zone in fracture mechanics. However,  due to the lack of permanent deformation upon stretching, we expect that the corner singularity to be regularized by nonlinear elasticity.

Finally, from the perspective of wrinkling in elastic sheets, we confirm with our experiments that a stretched sheet remains planar because the sheet thickness chosen in our study are above the critical threshold required for wrinkling~\cite{Nayyar2011,Healey2013}. In case where transverse wrinkles are observed in sheets with~\cite{kudrolli2018tension}, or without twist~\cite{Nayyar2011}, it was observed that the wrinkles form only in the central regions and are suppressed near the clamped boundaries to a distance which is of order of half the width of the sheet.  With our analysis of the clamp length $L_C$, we provide a quantitative characterization of the range over which the clamps induce a significant contribution to elastic field which has important implications to the development of models of tensional wrinkling.
According to our study, the stress fields in the corner dominated subregion extend from each corner to a similar distance. This biaxial stretching appears to suppress the formation of wrinkles in that region even when the sheet is thin enough to display transverse wrinkles. The observed extent of the corner dominated subregion may be further consistent with reports~\cite{Nayyar2011,Healey2013} that wrinkles are not observed for sheets with $L/W$ well below two, even in case of sufficiently thin sheets. Future work is required to connect the comprehensive analysis of the state of stress of the sheet provided by our study in the pre-wrinkling regime to the occurrence and extent of wrinkling in thin sheets.

\appendix
\section{Auxiliary functions for the asymptotic displacement fields}
\label{apend} 
The analytic expressions of the asymptotic displacement fields we use are adapted from Stern and Soni~\cite{stern1976computation}. There, fields were formulated using polar coordinates. Here, we provide the corresponding formulations using cartesian coordinates by introducing auxiliary functions %$g_x(\nu)$ and $ g_y(\nu)$ read:
\begin{align}
g_x(\nu) &= f_r^I(0) + \eta f_r^{II}(0)\,,\\
g_y(\nu) &= f_{\theta}^I(0) + \eta f_{\theta}^{II}(0)\,
\end{align}
where, $\eta = K_I/K_{II}$ is the mixed-mode ratio, $K_I$ and $K_{II}$ are the equivalent of the stress intensity factors and depend on the applied load and $\nu$. The mixed-mode ratio is given by
\begin{equation}
    \eta = \frac{\sin(\alpha \pi)}{\kappa+2\alpha+\cos(\alpha \pi)},
	\label{eq:eta}
 \end{equation}
where $\alpha$ is the exponent of the singularity which is given by Eq.~\ref{eq:lambda}. Eq.~\ref{eq:eta} implies that the local elastic field near the corner always includes an opening mode and a shear mode as commonly observed in interfacial crack between two different materials~\cite{Hutchinson1991}.
 
Then, the expressions for $\Phi_x(\theta)$ and $\Phi_y(\theta)$ reads:
\begin{align}
\Phi_x(\theta) = &g_x^{-1}(\nu)\left(f_r^I(\theta) + \eta \, f_r^{II}(\theta)\right)\,\cos \theta\, \\
- &g_x^{-1}(\nu)\left(f_{\theta}^I(\theta) + \eta \, f_{\theta}^{II}(\theta)\right)\,\sin \theta \,,\\
\Phi_y(\theta) = 
&g_y^{-1}(\nu)\left(f_r^I(\theta) + \eta \, f_r^{II}(\theta)\right)\,\sin \theta \\ +&g_y^{-1}(\nu)\left(f_{\theta}^I(\theta) + \eta \, f_{\theta}^{II}(\theta)\right)\,\cos \theta\,,
\end{align}
where,
\begin{align}
f_r^I (\theta)=&(\kappa-\alpha)  \cos[ (1-\alpha)(\pi/2-\theta)]\nonumber\\ 
-& (\kappa-a) \cos [(1+\alpha)(\pi/2-\theta)]\,,\\
f_{r}^{II} (\theta)=  -&(\kappa-\alpha) \sin[ (1-\alpha)(\pi/2-\theta)] \nonumber \\
+&(\kappa+a) \sin [(1+\alpha)(\pi/2-\theta)]\,,\\
f_{\theta}^{I} (\theta)=&  (\kappa+\alpha) \sin [(1-\alpha)(\pi/2-\theta)] \nonumber\\
-&(\kappa-a)  \sin[ (1+\alpha)(\pi/2-\theta)]\,,\\
f_{\theta}^{II} (\theta) =&  (\kappa+\alpha) \cos[ (1-\alpha)(\pi/2-\theta)] \nonumber\\
-&(\kappa+\alpha)  \cos [(1+\alpha)(\pi/2-\theta)]\,.
\end{align}

\begin{acknowledgments}
This work was supported by the National Science Foundation under grant number DMR 1508186.  
\end{acknowledgments}

\end{document}